\documentclass[aps,prl,superscriptaddress,showpacs,twocolumn]{revtex4}
\usepackage{graphicx}
\usepackage{amsfonts}
\usepackage{color}
\usepackage{amsmath, amsthm, amssymb}
\newcommand{\ie}{i.e.\ }
\newcommand{\eg}{e.g.\ } 

\begin{document}
\title{Quantum Monte Carlo study of a one-dimensional phase-fluctuating condensate in a harmonic 
trap }
\date{\today}
\author{C. Gils}
\affiliation{Institut f\"ur Theoretische Physik, ETH Z\"urich, CH-8093
  Z\"urich, Switzerland}
\author{L. Pollet}
\affiliation{Institut f\"ur Theoretische Physik, ETH Z\"urich, CH-8093
  Z\"urich, Switzerland}
\author{A. Vernier}
\author{F. Hebert}
\author{G.G. Batrouni}
\affiliation{Institute Non-Lin\'eaire de Nice, UMR CNRS 6618,  Universit\'e de
  Nice-Sophia Antipolis, 1361 route des Lucioles, F-06560 Valbonne, France}
\author{M. Troyer}
\affiliation{Institut f\"ur Theoretische Physik, ETH Z\"urich, CH-8093
  Z\"urich, Switzerland}
\begin{abstract}
We study numerically the low-temperature behavior of a one-dimensional Bose gas
trapped in an optical lattice.  For a sufficient number of particles
and weak repulsive interactions, we find a clear regime of
temperatures where density fluctuations are negligible but phase
fluctuations are considerable, \ie, a quasicondensate.  In the weakly
interacting limit, our results are in very good agreement with those
obtained using a mean-field approximation. 
In coupling regimes beyond the validity of mean-field approaches, a phase-fluctuating condensate also appears, but the phase-correlation properties are qualitatively different. It is shown that quantum depletion plays an important role.
\end{abstract}
\maketitle 

In a spatially homogeneous one-dimensional (1D) gas of bosons, spontaneous symmetry
breaking is excluded at all temperatures $T$
\cite{mermin_wagner,hohenberg,pitaevskii_stringari}.  Long-range order
is absent in this system due to the fluctuations of the phase of the
order parameter, as seen in the asymptotic decay of the equal-time
single-particle Green's function (algebraic for $T=0$ and exponential
for $T>0$) \cite{schwartz,haldane,kane_kadanoff,reatto_chester}.
However, analytical studies of trapped 1D bose gases with a fixed
particle number, $N$, reveal new phenomena. For example, the ground
state of a noninteracting 1D bose gas in a harmonic trap with
frequency $\omega$ becomes macroscopically populated below a
temperature $T_c\simeq N\hbar \omega /\ln(2N)$, \ie, a Bose-Einstein
condensate (BEC) exists in the finite system
\cite{ketterle_vandruten}.  Several mean-field studies indicate that
weak repulsive interactions introduce an additional effect, namely,
that fluctuations of the density are suppressed at low temperatures,
and a phase-fluctuating condensate, or quasicondensate, appears
\cite{petrov2d,petrov,mora_castin,luxat_griffin,bogoliubov,bouchoule}; for 
 the homogeneous system see \cite{kagan}.
Roughly speaking,
this is a regime in temperature where the distribution of particles in
space is given by a temperature-independent Thomas-Fermi profile,
while thermal fluctuations of the phase are present and lead to a
phase-coherence length that is smaller than the condensate cloud.
Only below a much lower temperature do phase fluctuations also become
negligible and phase coherence extends over the complete condensate
cloud.  A phase-fluctuating condensate emerges only in low dimensions
and has been experimentally observed in sufficiently anisotropic 3D
trapping geometries \cite{experiments}.

In this paper, we verify the existence of
a 1D quasicondensate, on trapped optical lattices, starting from a
microscopic approach whose validity is not restricted to certain
parameter regimes.  Using quantum-Monte Carlo simulations of the
Bose-Hubbard model, we investigate the properties of the
phase-fluctuating condensate for various choices of interaction
strength and density.  For weak interactions, our results are in
excellent agreement with mean-field estimates in the
continuum~\cite{petrov}.  In an intermediate-coupling regime beyond
the weakly interacting limit, but not yet in the strong-coupling
domain, the quasicondensate regime spreads over an even larger
temperature range.  Nonetheless, we observe qualitatively different phase-correlation properties which do not follow from existing analytical
approaches.

First, we briefly review the notion of a quasicondensate (QC) in a
weakly interacting (WI) 1D trapped bose gas as presented in
\cite{petrov}.  In the WI limit, we have $\gamma = mg/\hbar^2n \ll 1
$, where $n$ is the average particle density, $m$ the particle mass,
and $g$ the coupling constant of a repulsive contact interaction
\cite{lieb_liniger}.  In second-quantized representation, the
Hamiltonian of the system is given by
\begin{equation}
\hat{H}= \int dz \,\hat{\psi}^{\dagger}(z) \left(
  -\frac{\hbar^2\nabla^2}{2m} + V_{\rm t}(z) + \frac{g}{2}
  \hat{\psi}^{\dagger}(z)\hat{\psi} (z)  \right ) \hat{\psi} (z),
\label{ham1}
\tag{1}
\end{equation}
where $\hat{\psi}(z)$ is the bosonic field operator and $V_{\rm t}(z)=
m\omega^2 z^2 /2$ the external trapping potential centered at the
origin.  The authors in \cite{petrov} show that density fluctuations
$\langle \delta \hat{n}(z) \delta \hat{n}(z') \rangle $, where
$\hat{n}(z)= n(z) + \delta \hat{n}(z)$, are suppressed for inverse
temperatures $\beta \gg \beta_d=(N\hbar \omega)^{-1}$ (Boltzmann
constant $k_B=1)$. The field operator can then be expressed as
$\hat{\psi}(z)=\sqrt{n(z)}\exp[i \hat{\phi}(z)]$.  Thermal
fluctuations of the phase are evident from the decay of the
one-particle density matrix, or (equal-time) Green's function, which
is obtained as \cite{petrov}
\begin{equation*}
\langle \hat{\psi}^{\dagger}(z) \hat{\psi}(z') \rangle = \sqrt{n(z)
n(z')} \exp (-\langle \delta \hat{\phi}^2_{zz'} \rangle/2),
\label{green}
\tag{2a}
\end{equation*}
where $\delta \hat{\phi}^2_{zz'} = (\hat{\phi}(z)-\hat{\phi}(z'))^2$,
and 
\begin{equation*}
 \langle \delta \hat{\phi}^2_{zz'} \rangle = \frac{4 \beta_d \mu_{\rm
 TF} }{3 \beta \hbar\omega} \left | \ln\left ( \frac{(L_{\rm
 TF}-z')}{(L_{\rm TF}+z')}\frac{(L_{\rm TF}+z)}{(L_{\rm TF}-z)}
 \right) \right |,
\label{phasefluct}
\tag{2b}
\end{equation*}
where $\mu_{\rm TF}=(3Ng/4)^{2/3}(m \omega^2/2)^{1/3}$ is the chemical
potential, and $L_{\rm TF} = \sqrt{2\mu_{\rm TF}/m\omega^2}$ the
half size of the condensate in the Thomas-Fermi (TF) approximation
($\mu_{\rm TF} \gg \hbar \omega$; see \eg \cite{bec_review}).  For
$\beta\to \infty$, it follows from Eq. (\ref{phasefluct}) that
$\langle \delta \hat{\phi}^2_{zz'}\rangle \to 0$, and thus $G(z,z')\to
\sqrt{n(z)n(z')}$, \ie the system is completely phase coherent.

As is well known \cite{jaksch}, the discretization of Eq.\ (\ref{ham1})
yields the Bose-Hubbard model with lattice Hamiltonian
\begin{align*}
\nonumber \hat{H} &=& - J \sum_{\langle j,j' \rangle} (
\hat{a}_j^{\dagger} \hat{a}_{j'} + {\rm H.c} ) + \frac{U}{2} \sum_j
\hat{n}_j (\hat{n}_j-1) \\ && \qquad +V_t \sum_j (j-M/2)^2 \hat{n}_j,
\label{ham2}
\tag{3}
\end{align*}
where $\hat{a}_j^{\dagger}$ creates a particle at optical lattice site
$j$ and $\hat{n}_j = \hat{a}_j^{\dagger} \hat{a}_j $. The number of
lattice sites $M$ is chosen such that the occupation at the boundaries
of the lattice is zero.  We work in units where the lattice spacing
and $\hbar^2/2m $ are set to 1. In these units, parameters in Eqs.\
(\ref{ham1}) and (\ref{ham2}) are related as follows: $J=1$, $U=g$,
$V_{\rm t}=(\hbar \omega)^2/4$.  Our system is defined on an optical
lattice, however, in the WI and degenerate limits, the discrete and
continuum description of the phase-fluctuating condensate are
equivalent \cite{mora_castin}.
\begin{figure}[t]
\begin{center}
\includegraphics[width=8cm,clip]{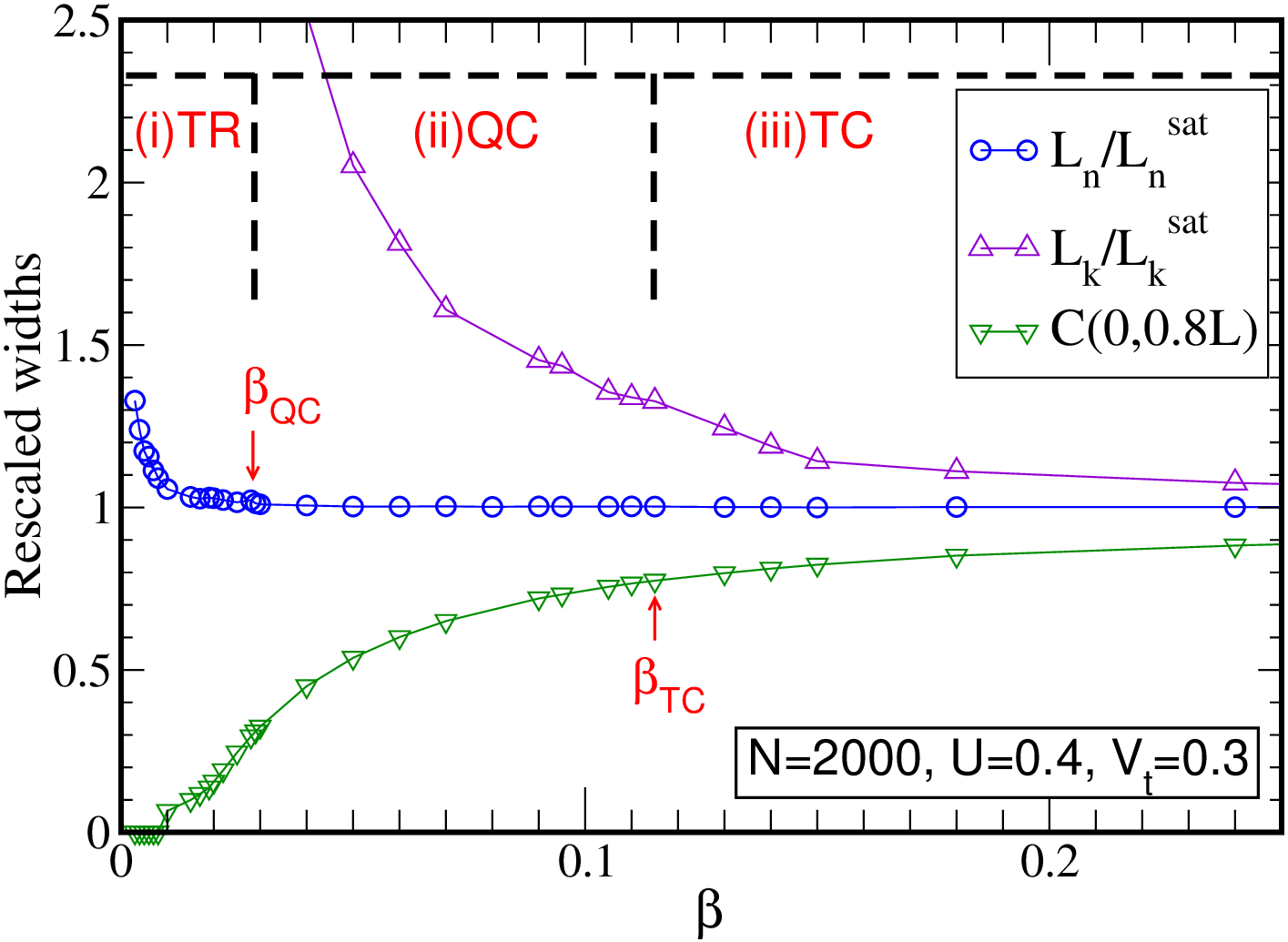}
\caption{(Color online)(i) Thermal regime: As a result of both density and phase
  fluctuations, the widths of the density profile, $L_n$, and the
  momentum profile, $L_k$, depend on $\beta$ with $L_n(\beta)
  >L_n^{\rm sat}$, $L_k(\beta) > L_k^{\rm sat}$.  (ii)
  Quasicondensate: Density fluctuations are negligible ($L_n=L_n^{\rm
  sat}$, $L$ is the half size of the condensate cloud for $\beta\ge
  \beta_{\rm QC}$). However, the phase fluctuates, as seen for  $L_k(\beta)>L_k^{\rm sat}$, phase correlation function
  $C(0,0.8L)<1$.  (iii) True condensate: Phase fluctuations also
  disappear as $L_k(\beta)$ approaches $L_k^{\rm sat}$, and
  $C(0,0.8L)$ approaches 1.  The numerical results are $\beta_{\rm
  QC}\approx 0.03$ and $\beta_{\rm TC}=0.115(5)$ (analytical estimate:
  $\tilde{\beta}_{\rm TC}=0.116$).}
\label{LvsBeta}
\end{center}
\end{figure}
We investigate this model using a worm update quantum Monte Carlo
(QMC) method in the canonical ensemble \cite{can_worm}. This nonlocal
update scheme allows for an efficient evaluation of the equal-time
Green's function $G(j,j') = \langle \hat{a}_j^{\dagger}\hat{a}_{j'}
\rangle$. Phase correlation properties are also apparent from the
shapes of the momentum profile, $n(k) = \sum_{j,j'} G(j,j')
\exp[ik(j-j')]$, and the rescaled Green's function (phase correlation
function), $C(j,j') = G(j,j')/\sqrt{n_j n_{j'}}$, where $n_j=\langle
\hat{n}_j \rangle$ is the density distribution.  In the WI, TF and
mean-field ($N$ large enough) limits, we observe three different
regimes in temperature: the thermal regime (TR), the quasicondensate
(QC) and the ``true condensate'' (TC). These regimes are separated by
smooth crossovers, which are characterized by the temperatures
$1/\beta_{\rm QC}$ and $1/\beta_{\rm TC}$.  We discuss the properties
of the three regimes and compare our results with the mean-field
results listed above.  Furthermore, we consider the emergence of the
phase-fluctuating condensate depending on the choice of parameters
$N$, $U$ and $V_t$, including parameter sets that are beyond the WI
limit.
 
\begin{figure}[t]
\begin{center}
\includegraphics[width=8cm,clip]{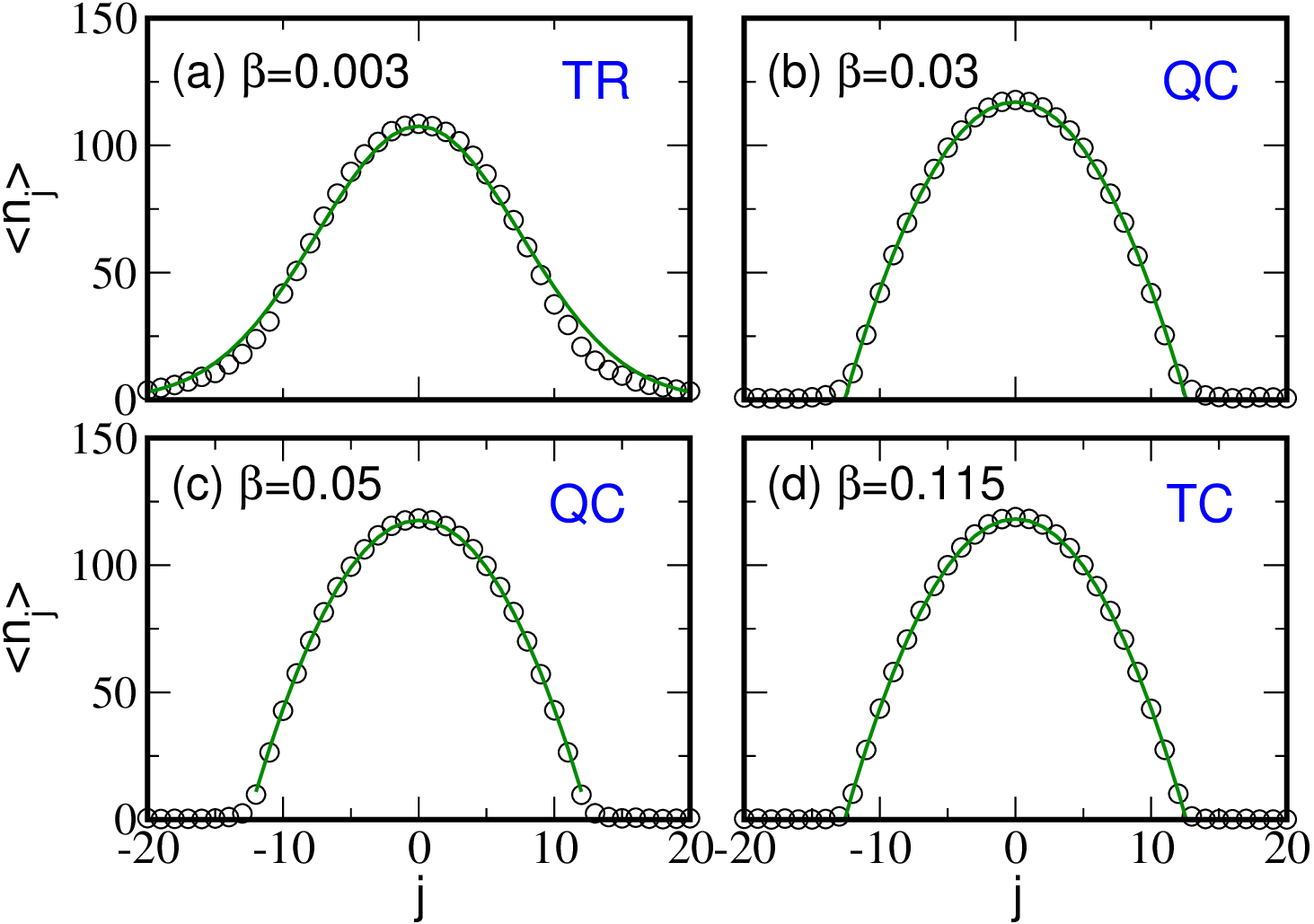}
\caption{(Color online) $N=2000$, $U=0.4$, $V_t= 0.3$ ($\gamma=0.003$). Density
  fluctuations are suppressed for $\beta\ge \beta_{\rm QC}\approx
  0.03$ and the density profiles $n_j=\langle \hat{n}_j \rangle$
  ($\circ$) are identical for $\beta\ge \beta_{\rm QC}$ (b)-(d). In
  (a), the line is a fit to a Gaussian profile, while in (b)-(d) the
  lines are fits to a TF profile $n_0 (1-j^2/L^2)$ where $L=12.6$
  ($L_{\rm TF} = 12.6$).  Error bars in the figures are always smaller
  than the symbol size.}
\label{denprofiles}
\end{center}
\end{figure}

\begin{figure}
\begin{center}
\includegraphics[width=8.5cm,clip]{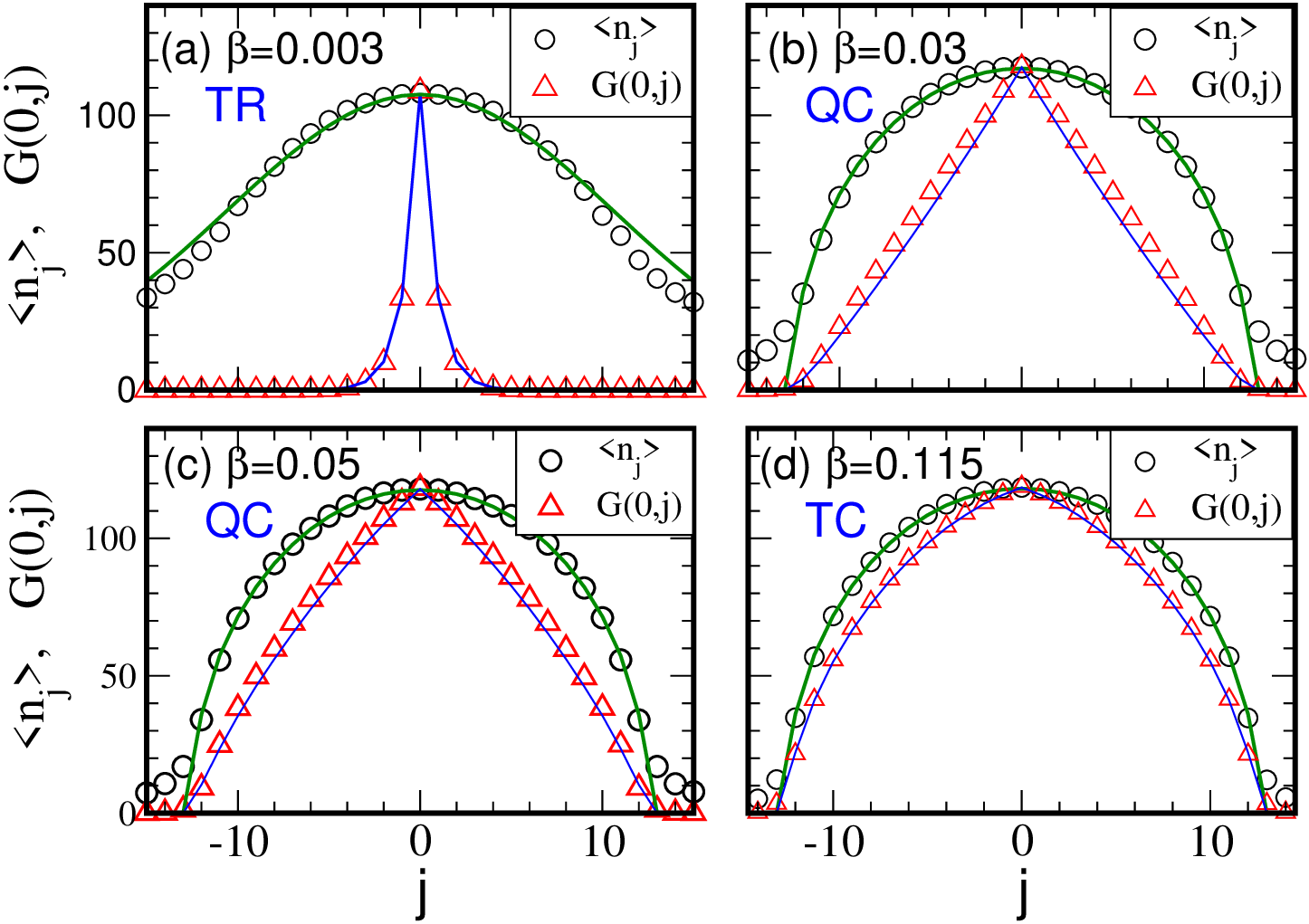}
\caption{(Color online) $N=2000$, $U=0.4$, $V_t= 0.3$.  The density profiles
$\sqrt{n_0n_j}$ ($\circ$) are virtually independent of temperature for
$\beta \ge \beta_{\rm QC}\approx 0.03$ (b)-(d), while the Green's
function $G(0,j)$ ($\triangle$) varies with $\beta$.  The lines are
the analytical estimates, a Gaussian fit for $\sqrt{n_0 n_j}$ in (a), a TF
profile $n_0\sqrt{1-j^2/L^2}$ for $\sqrt{n_0n_j}$ in (b)-(d) and 
$\sqrt{n_0n_j}\exp(-\alpha |\ln[(L-j)/(L+j)]|)$ [see Eq.\ (2)] for $G(0,j)$.  }
\label{profile-green}
\end{center}
\end{figure}

In the thermal regime $\beta < \beta_{\rm QC}$, both
the density and the phase are governed by thermal fluctuations. Hence,
all quantities exhibit a strong temperature dependance. The width of
the density profile, $L_n(\beta)$ (standard deviation of $n_j$),
decreases throughout the TR, until it reaches its minimum value
$L_n^{\rm sat}$ at the inverse temperature $\beta_{\rm QC}$, as shown
in Fig.~\ref{LvsBeta}.  The shape of the density profile is
approximately Gaussian, which is expected for high temperatures where
the bosonic nature of the particles becomes less relevant
[Fig.~\ref{denprofiles} (a)].  The strong phase fluctuations are seen
in the width of the momentum profile, $L_k(\beta)$ [standard deviation
of $n(k)$], which is much larger than its minimum value $L_k^{\rm
sat}$ [Fig.~\ref{LvsBeta}], as well as in the exponential decay of the
Green's function [Fig.~\ref{profile-green} (a)].  We recall that there
exists no analytical estimate for $\beta_{\rm QC}$; density
fluctutations are merely predicted to become small for $\beta \gg
\beta_d=(4V_tN^2)^{-1/2}$ \cite{petrov}.  For the parameter set in the
Figs.~\ref{LvsBeta}-\ref{profile-green} , we have $\beta_d \approx
0.02 \beta_{\rm QC}$.

In the quasicondensate ($\beta_{\rm QC}\le\beta < \beta_{\rm TC}$),
the density no longer fluctuates: the density profiles $n_j$ at
different temperatures $\beta\ge\beta_{\rm QC}$ are identical and in
excellent agreement with a TF inverse parabola shape of half size $L$,
where $L$ equals to $L_{\rm TF}$ [Figs.~\ref{denprofiles}(b)-~\ref{denprofiles}(d)].
However, thermal fluctuations of the phase are considerable. Thus, the
width of the momentum profile, and the phase correlation function
still change with decreasing temperature
(Fig.~\ref{LvsBeta}). Figures~\ref{profile-green}(b)-~\ref{profile-green}(d)
demonstrate that the Green's function varies substantially for different $\beta\ge
\beta_{\rm QC}$, while the density profile $\sqrt{n_0 n_j}$ is
invariant.  We find very good agreement with the mean-field result
Eq. (2): the ansatz $\sqrt{n_0n_j}\exp\{-\alpha |\ln[(L-j)/(L+j)]|\}$,
with $\alpha$ a fitting parameter, reproduces $G(0,j)$
(Fig.~\ref{profile-green}).

  \begin{figure}[t]
\includegraphics[width=8cm,height=7.5cm,clip]{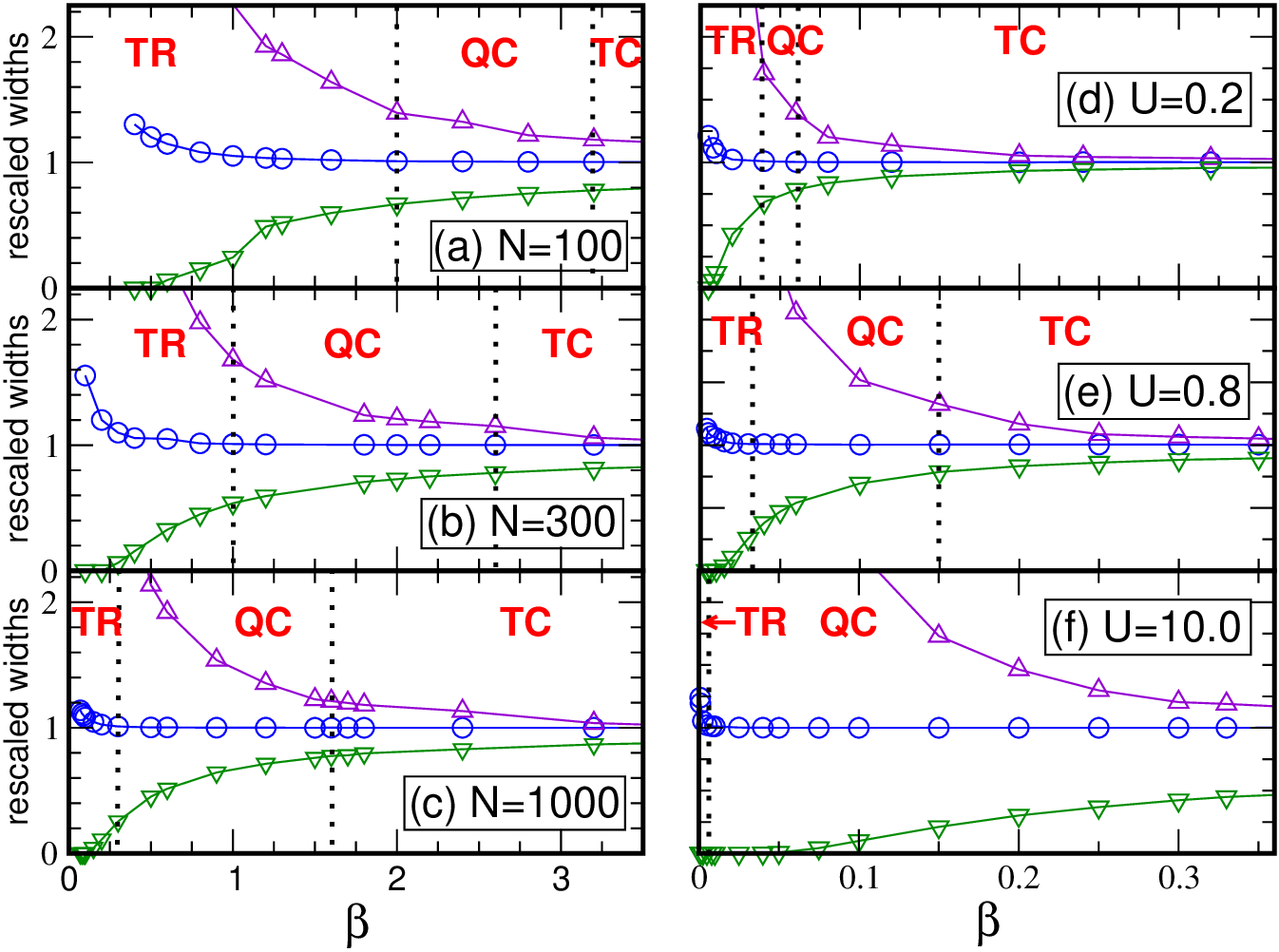}
\caption{(Color online) The size $s=\beta_{\rm TC}/\beta_{\rm QC}$, of the QC regime
  depending on the number of particles $N$ (a - c, $U=0.5$,
  $V_t=0.01$) and coupling constant $U$ (d - f, $N=2000$,
  $V_t=0.4$). \\ $L_n/L_n^{\rm sat}$ ($\circ$), $L_k/L_k^{\rm sat}$
  ($\bigtriangleup$), $C(0,0.8L)$ ($\bigtriangledown$) \\  a)
  $\gamma=0.08$, $\tilde{\beta}_{\rm TC}=3.5$, $\beta_{\rm
    TC}\approx 3.2(4),\beta_{\rm QC}\approx 2.0$, $s \approx 1.6$\\ b)
  $\gamma=0.04$, $\tilde{\beta}_{\rm TC}=2.5$, $\beta_{\rm
    TC}\approx 2.6(4),\beta_{\rm QC}\approx 1.0$, $s \approx 2.6$\\  c)
  $\gamma=0.02$, $\tilde{\beta}_{\rm TC}=1.64 $, $\beta_{\rm
  TC} \approx 1.6(1),\beta_{\rm QC}\approx 0.3$ , $s\approx 5.3$\\  d) $\gamma=0.001$, $\tilde{\beta}_{\rm 
TC}=0.06$, $\beta_{\rm
  TC} \approx 0.06(2),\beta_{\rm QC}\approx 0.04 $, $s\approx 1.5$\\  e) $\gamma=0.006$, $\tilde{\beta}_{\rm 
TC}=0.15$, $\beta_{\rm
  TC}\approx 0.15(5),\beta_{\rm QC}\approx 0.03 $, $s\approx 5$\\ f)
  $\gamma=0.17$, $\tilde{\beta}_{\rm TC}=0.82 $, NO $\beta_{\rm TC}$,
  $\beta_{\rm QC}\approx 0.01 $, $s \approx \beta_{\rm QC}^{-1}=100
  $ \\Comparison of (a)-(c), and (d)-(f), respectively, shows that $s$
  increases with increasing $N$ and $U$.  }
\label{profile_decay_2}
\end{figure}

At temperatures much lower than $1/\beta_{\rm QC}$, phase fluctuations
also become suppressed. Therefore, it is meaningful to introduce a
second crossover temperature, $1/\beta_{\rm TC}$, to a phase-coherent
regime, the true condensate.  We define $\beta_{\rm TC}$ as the
temperature where $C(0,0.8L)= \exp(-0.25)$.  This definition is
somewhat arbitrary, but yields a scale below which the system can
safely be considered to be phase-coherent.  In addition, it allows for a
comparison of analytical and numerical results, with the analytical
equivalent of $\beta_{\rm TC}$ being $\tilde{\beta}_{\rm TC} =
4\beta_d \mu_{\rm TF}\ln(9)/ 3\hbar\omega$ [which is the temperature
where $\langle \delta \hat{\phi}^2_{0z'} \rangle= 0.5$ for $|z'|=0.8
L_{\rm TF}$; see Eq.\ 2]. Note that the true condensate is not to be
confused with a BEC, and strictly only appears at zero temperature
(Thomas-Fermi condensate; see \cite{petrov}).  In the true
condensate, the Green's function and the momentum profile approach
their temperature-independent ground state shapes, as shown by the
convergence of $L_k(\beta)$ and $C(0,0.8L)$ in Fig.~\ref{LvsBeta}.
The system becomes practically phase coherencent, as illustrated in
Fig.~\ref{profile-green}(d), where $\sqrt{n_0 n_j} \approx G(0,j)$.
For the parameter set in Figs.~\ref{LvsBeta}-\ref{profile-green}, we find that $\beta_{\rm
TC} \approx 0.115(5)$ which agrees with the analytical estimate
$\tilde{\beta}_{\rm TC}=[(U^2/6NV_t^2)^{1/3}\ln(9)]=0.116$.

We now study the appearance of a phase-fluctuating condensate
depending on the system parameters $U$, $V_t$, and $N$.  The WI limit
is characterized by $\gamma = U / 2n\ll 1$. We define the average density by $n=N/2L$. Since we
find that $L$ equals $L_{\rm TF}$ within error bars in all cases, we
use $\gamma = (3U^4/4V_tN^2)^{1/3}$. Note that the magnitude of
$\gamma$ in the inhomogenous system differs from that in the
homogeneous system. 
 In Figs.~\ref{profile_decay_2}(a)-~\ref{profile_decay_2}(c), we
demonstrate the effect of varying the number of particles $N$ at
on-site repulsion $U=0.5$ and trapping $V_t=0.01$. Both $\beta_{\rm
QC}$ and $\beta_{\rm TC}$ decrease with increasing $N$, but since the
change of $\beta_{\rm QC}$ is much greater than that of $ \beta_{\rm
TC}$, the region of the phase-fluctuating condensate, $s= \beta_{\rm
TC}/\beta_{\rm QC}$, increases with increasing $N$.  The effect of
varying $U$ is illustrated in Figs.~\ref{profile_decay_2}(d)-~\ref{profile_decay_2}(f),
where $N=2000$ and $V_t=0.4$. Increasing $U$, causes $\beta_{\rm QC}$
to decrease, while $\beta_{\rm TC}$ increases. Thus $\beta_{\rm
TC}/\beta_{\rm QC}$ grows with increasing $U$.
More generally, the deeper we are in the Thomas-Fermi limit of large $\mu_{\rm TF}/\hbar \omega \sim [(NU)^4/V_t]^{1/6}$,
 the larger is the size of the quasicondensate.

\begin{figure}[t]
\includegraphics[width=8.5cm,clip]{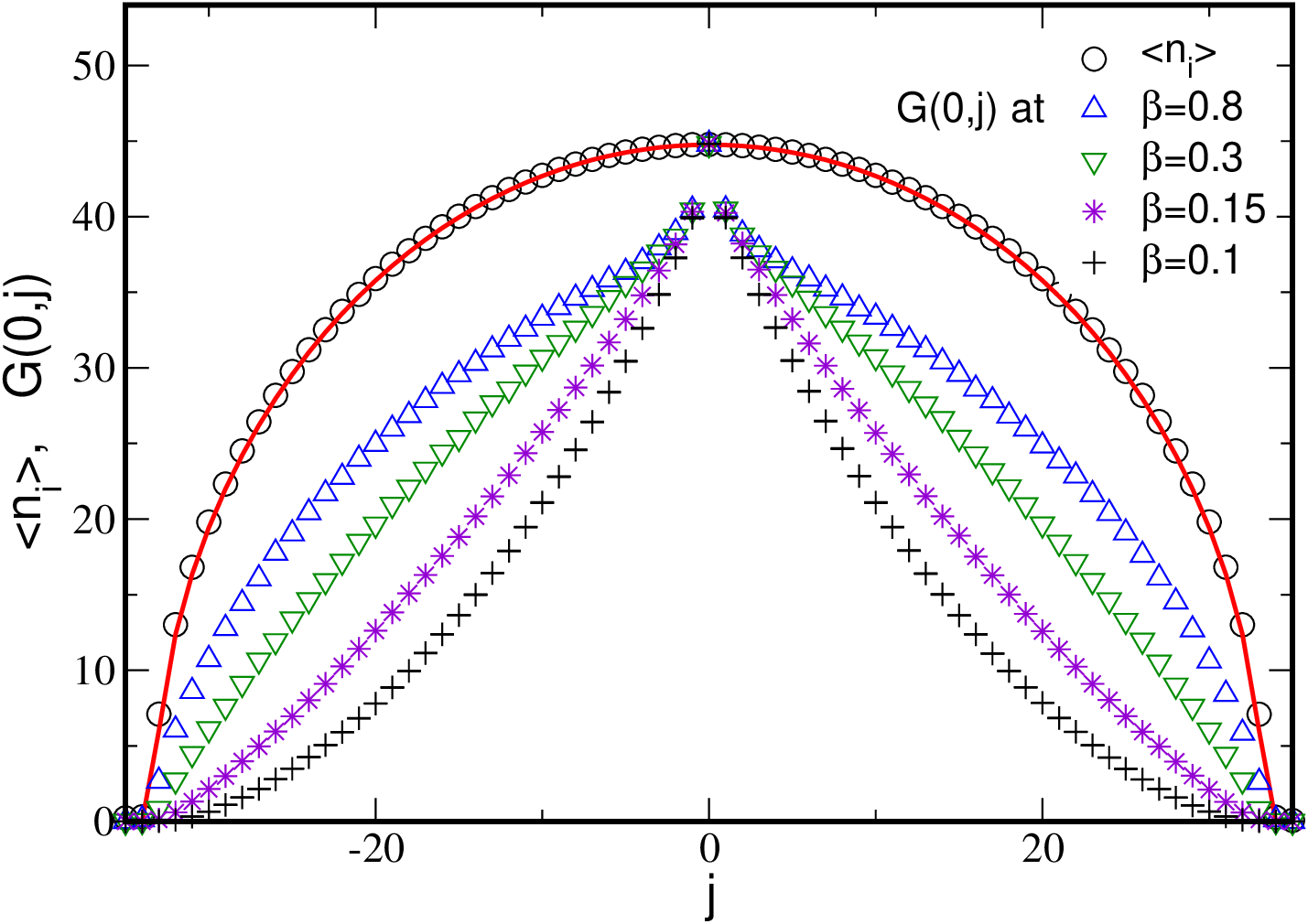}
\caption{(Color online) $N=2000$, $U=10.0$, $V_t=0.4$ ($\gamma=0.17$).  The density
profile $\sqrt{n_0n_j}$ ($\circ$) is invariant for $\beta \ge
\beta_{\rm QC}\approx 0.1$, while the Green's function $G(0,j)$ varies
throughout the QC regime. However, $G(0,j)$ does not approach
$\sqrt{n_0n_j}$ (as in Fig.~\ref{profile-green}).  The line is a TF
fit to $\sqrt{n_0 n_j}$ with $L=33.3$
($L_{\rm TF}=33.5$). }
\label{profile-green_2}
\end{figure}

While for the parameter sets in Figs.~\ref{profile_decay_2}(a)-~\ref{profile_decay_2}(e),
good agreement of $G(0,j)$ with expression (2) is observed for all
$\beta$, as well as $\beta_{\rm TC}\approx \tilde{\beta}_{\rm TC}$,
this does not apply to the example in Fig.~\ref{profile_decay_2}(f),
where $\gamma=0.17$.  Instead, $G(0,j)$ exhibits a qualitatively
different behavior when approaching the ground state, as can be seen
in Fig.~\ref{profile-green_2}: the phase-fluctutating condensate is
beautifully realized, however, $G(0,j)$ does not approach $\sqrt{n_0
n_j}$ for $T \to 0$, and cannot be described by Eq.\ (2).


\begin{figure}[t]
\includegraphics[width=8.5cm,clip]{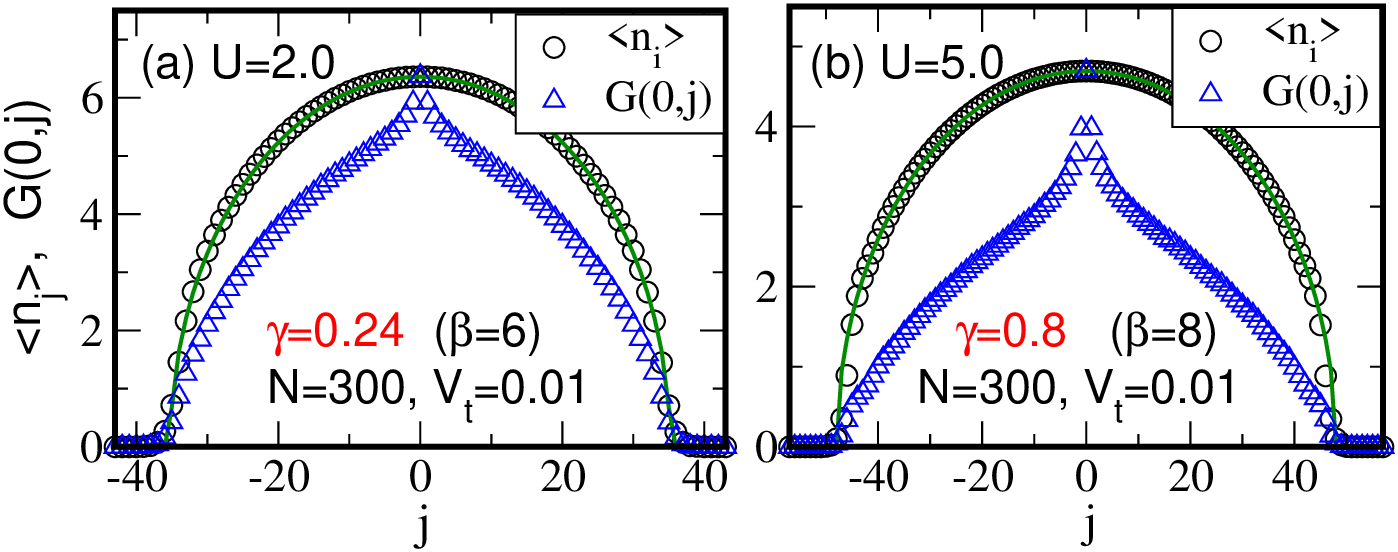}
\caption{(Color online) Ground state $\sqrt{n_0n_j}$ ($\circ$) and $G(0,j)$
($\triangle$) for different  $\gamma$. The lines are fits to a TF profile with
  (a) $L=35.2$ ($L_{\rm TF} = 35.6$) and (b) $L=48.0$ ($L_{\rm TF}=48.3$). For increasing $\gamma$, the correlation hole becomes more
pronounced, and thus the quantum depletion increases.}
\label{fig6}
\end{figure}

We consider the effect of stronger interactions in the trapped system
on the phase-correlation properties in more detail.  In
Fig.~\ref{fig6}, we show the ground state profiles for $N=300$,
$V_t=0.01$, $U=2.0$ and $U=5.0$, respectively.  We observe the same
qualitative behaviour of the saturated Green's function as in
Fig.~\ref{profile-green_2}.  Close to the center of the cloud, where
the density is higher, the decay of $G(0,j)$ is exponential; however,
it broadens toward the outer, more diluted regions of the cloud. 
By
comparing Figs.~\ref{fig6}(a) and ~\ref{fig6}(b), it can be seen that for
larger $\gamma$, the correlation hole in the central region of the
condensate is larger.
Clearly, the quantum depletion for the parameter sets in Fig.~\ref{fig6} is
much more significant than  for systems  in the WI limit,
since the number of particles with zero momentum 
$ N(k=0) = \sum_{j,j'} G(j,j')/M$ decreases  if $G(j,j') < \sqrt{n_j n_{j'}}$.  
We also observe this behavior if the density
is smaller than 1 everywhere in the trap, and therefore exclude the
possibility that it is an effect of the optical lattice.  The shape of $G(0,j)$ is
consistent with the two limiting cases, \ie exponential decay in the
strong-coupling limit, and phase-coherence [\ie Eq.(2)] in the
weak-coupling limit.

We conclude with a summary of our main results.  In the mean-field,
weakly interacting and Thomas-Fermi limits, both true a Thomas-Fermi
condensate and a phase-fluctuating condensate emerge.  Phase correlation
properties, manifest in the characteristic decay of the
single-particle density matrix, agree with surprising precision with
the mean-field theory in \cite{petrov}.  In an intermediate-coupling
regime, a true condensate no longer appears. The regime of the
phase-fluctuating condensate persists even longer, extending to $T=0$.
We observe a qualitatively different decay of the Green's function,
which cannot be accounted for by mean-field studies.

We acknowledge helpful discussions with Y. Castin, G. Shlyapnikov and
T. Esslinger. This work was supported by a CNRS PICS grant (G.G.B. and F.H.) and
the Swiss National Science Foundation. The simulations were performed on the Hreidar Beowulf cluster at ETH Zurich.


\end{document}